\documentclass{article}
\usepackage[accsupp]{axessibility}  

\usepackage{ijcai23}

\usepackage{times}
\usepackage{soul}
\usepackage{url}
\usepackage[hidelinks]{hyperref}
\usepackage[utf8]{inputenc}
\usepackage[small]{caption}
\usepackage{graphicx}
\usepackage{amsmath}
\usepackage{amsthm}
\usepackage{booktabs}
\usepackage{algorithm}
\usepackage{algorithmic}
\usepackage[switch]{lineno}
\usepackage{subfigure}  
\usepackage{graphicx}
\usepackage{amssymb}

\usepackage[labeled]{multibib}
\newcites{A}{Appendix References}
\usepackage[misc]{ifsym}

\title{DiffuseStyleGesture: Stylized Audio-Driven Co-Speech Gesture Generation with Diffusion Models}

\author{
Sicheng Yang$^1$\and
Zhiyong Wu$^{1,4*}$\and       
Minglei Li$^2$\and
Zhensong Zhang$^3$\and \\
Lei Hao$^3$\and
Weihong Bao$^1$\and
Ming Cheng$^1$\and
Long Xiao$^1$
\affiliations
$^1$Shenzhen International Graduate School, Tsinghua University, Shenzhen, China\\
$^2$Huawei Cloud Computing Technologies Co., Ltd, Shenzhen, China\\
$^3$Huawei Noah's Ark Lab, Shenzhen, China\\
$^4$The Chinese University of Hong Kong, Hong Kong SAR, China
\emails
yangsc21@mails.tsinghua.edu.cn,
zywu@sz.tsinghua.edu.cn, \\
\{liminglei29, zhangzhensong\}@huawei.com
}

\begin{document}

\maketitle


\begin{abstract}
    The art of communication beyond speech there are gestures. The automatic co-speech gesture generation draws much attention in computer animation. It is a challenging task due to the diversity of gestures and the difficulty of matching the rhythm and semantics of the gesture to the corresponding speech.
    To address these problems, we present \textbf{DiffuseStyleGesture}, a diffusion model-based speech-driven gesture generation approach. It generates high-quality, speech-matched, stylized, and diverse co-speech gestures based on given speeches of arbitrary length.
    Specifically, we introduce cross-local attention and self-attention to the gesture diffusion pipeline to generate better speech-matched and realistic gestures.
    We then train our model with classifier-free guidance to control the gesture style by interpolation or extrapolation. Additionally, we improve the diversity of generated gestures with different initial gestures and noise.
    Extensive experiments show that our method outperforms recent approaches on speech-driven gesture generation.
    Our code, pre-trained models, and demos are available at \url{https://github.com/YoungSeng/DiffuseStyleGesture}.
\end{abstract}


\section{Introduction}      

\begin{figure}[!ht]
  \centering
   \includegraphics[width=\linewidth]{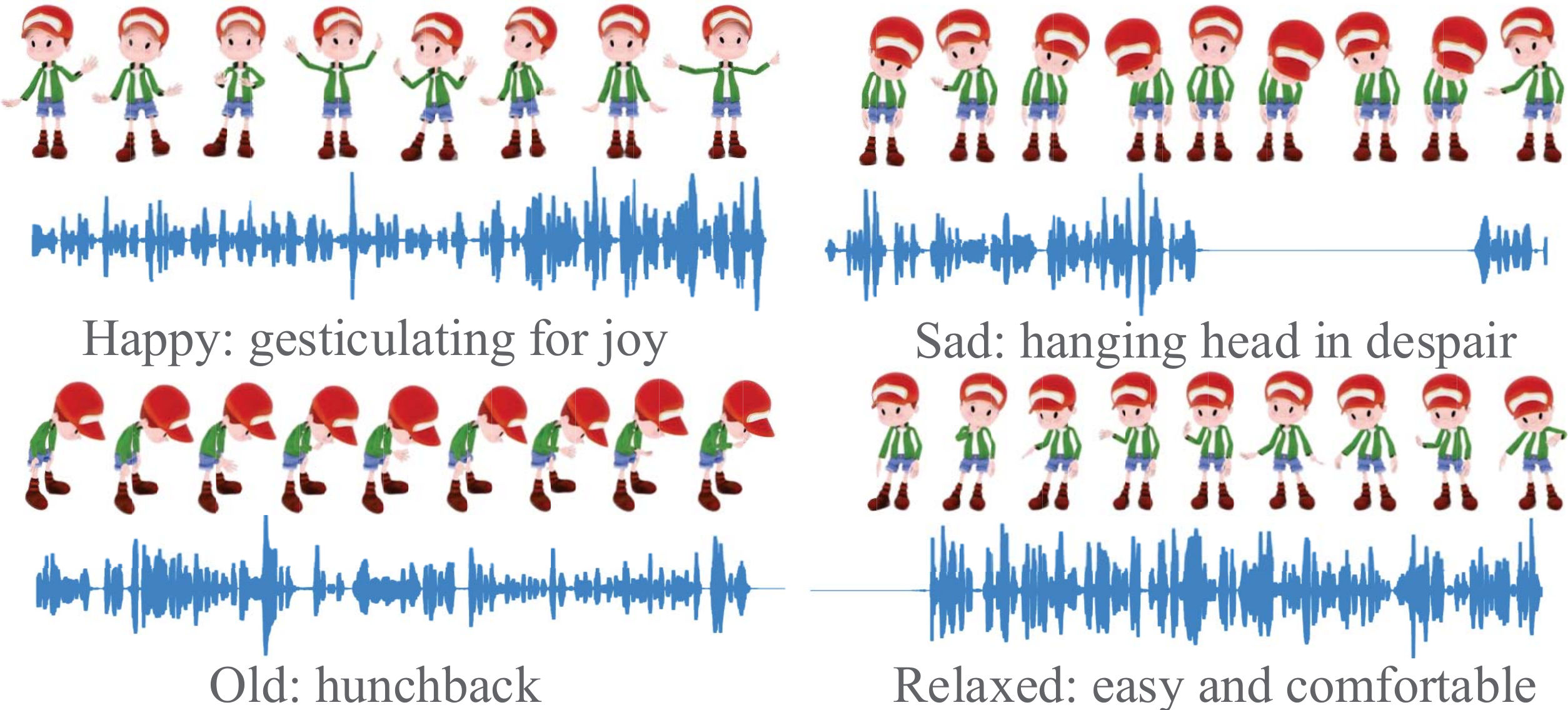}
   \caption{\textbf{Gesture examples generated by our proposed method} on various types of speech and styles. 
  All characters used in the paper are publicly available.}
   \label{fig:overview}
\end{figure}

Body gestures and facial expressions are important tools for conveying information in human communication \cite{DBLP:conf/iui/KucherenkoJYWH21}.
Automated generation of co-speech gestures is a crucial technology for developing lifelike avatars in movies, gaming, virtual social environments, and interactions with social robots \cite{nyatsanga2023comprehensive}.
The most important issues of co-speech gesture generation are 1) how to generate gestures matching the rhythm of audio and semantics of text; and 2) how to generate diverse and stylized gestures.
Recent gesture generation methods can directly generate human gestures conditioned on neutral speech \cite{nyatsanga2023comprehensive,DBLP:conf/icmi/YoonWKVNTH22,DBLP:conf/iui/KucherenkoJYWH21}.
However, all these approaches still limit the learned distribution since they mainly employ GAN-based \cite{yoon2020speech},    
VAEs \cite{DBLP:conf/iccv/0071KPZZ0B21}     
or Flows \cite{DBLP:journals/cgf/AlexandersonHKB20}.        
GAN-based synthesis methods suffer from mode collapse, which leads to low-quality synthesis, especially with data unseen in the training data. 
Methods using VAEs and Flows require a trade-off between generation quality and diversity \cite{tevet2022human,DBLP:journals/corr/abs-2212-04495}.

Recently, diffusion models \cite{ho2020denoising}      
which are generative approaches have achieved impressive results in other domains due to their high quality and diversity of generation, such as image generation \cite{ramesh2022hierarchical},     
video generation \cite{DBLP:journals/corr/abs-2212-00235},        
and text generation \cite{DBLP:journals/corr/abs-2212-11685}.     
These works demonstrate the ability of denoising-diffusion-based models to learn real data distributions while also providing diverse sampling and manipulation, such as editing and interpolation.
However, these works do not model timing-dependent sequences to solve temporal aligned problems like speech-driven gestures, and they are computationally resource-intensive.

\begin{figure*}[!ht]
  \centering
   \includegraphics[width=0.85125\linewidth]{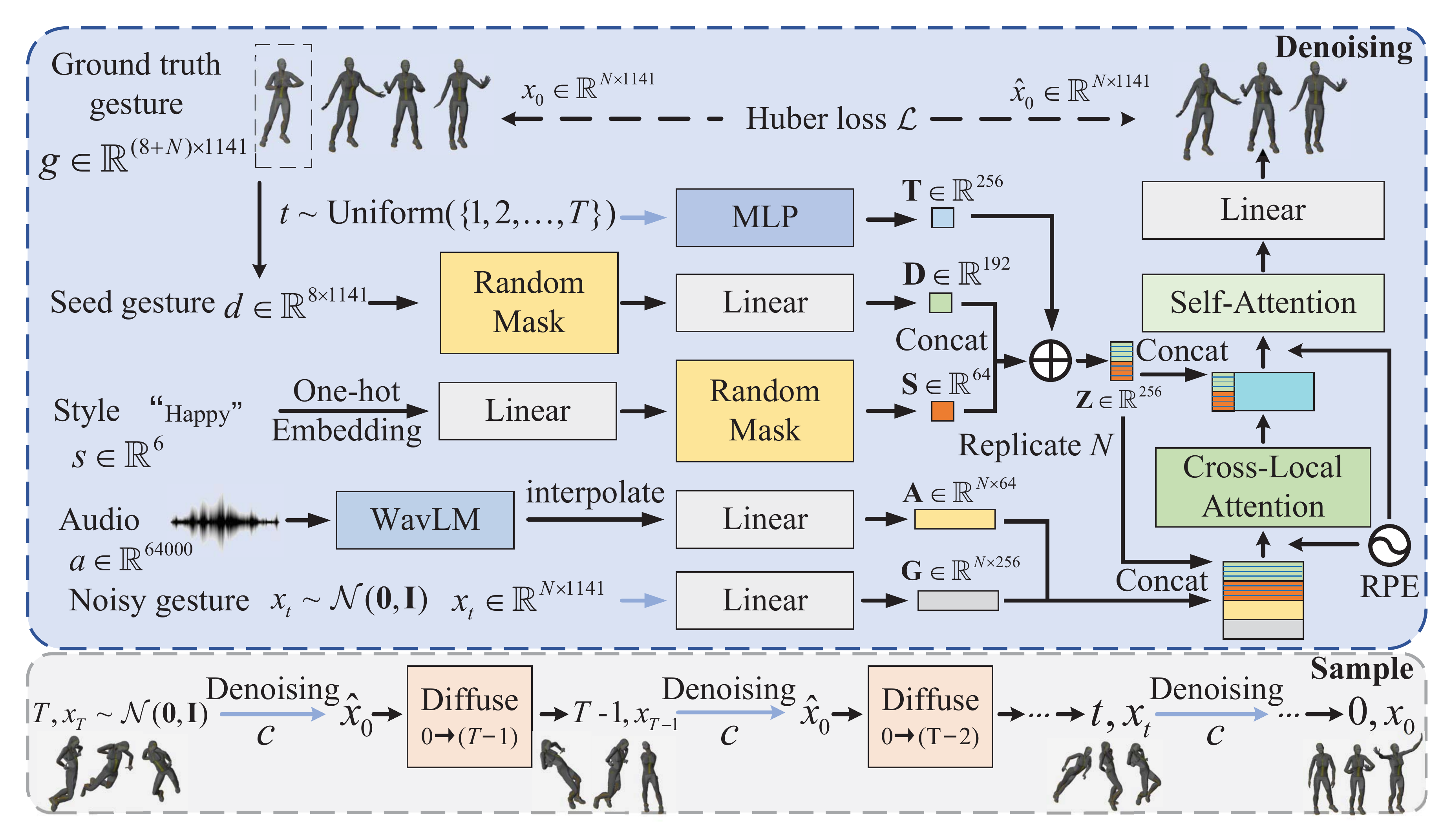}
   \caption{\textbf{(Top) Denoising module of DiffuseStyleGesture.}
   A noising step $t$ and a noisy gesture sequence $x_t$ at this noising step conditioning on $c$ (including seed gesture $d$, style $s$, and audio $a$) are fed into the model.
   Cross-local attention and self-attention can better capture the correlations between speech and gesture based on WavLM features.
   Random masks in the seed gesture and style feature processing pipeline help classifier-free guidance training of the model and perform interpolation or extrapolation to achieve a high degree of control over the generated gestures.
   \textbf{(Bottom) Sample module of DiffuseStyleGesture.}
   At each step $t$, we predict the $\hat{x}_0$ with the denoising process based on the corresponding conditions, then add the noise to the noising step $x_{t-1}$ with the diffuse process.
   This process is repeated from $t$ = $T$ until $t=0$.
   }
   \label{fig:Framework}
\end{figure*}

To generate high-quality, speech-matched, stylized, and diverse co-speech gestures, inspired by the recent progress of the denoising-diffusion-based generation, we introduce \textbf{DiffuseStyleGesture}, a versatile, controllable, and time-aware denoising-diffusion-based model for audio-driven co-speech gesture generation.
Examples of the generated gesture are shown in Figure \ref{fig:overview}.
And the overview of our method is shown in Figure \ref{fig:Framework}.
We use an attention-based architecture to capture the temporal information between speech and gestures.
And we find that it is better to train the model to predict the signal itself \cite{ramesh2022hierarchical,tevet2022human} than to predict the noise.
To align the generated gestures better with the speech, we also propose an approach that uses cross-local attention to capture local information of gestures and speech, and then uses self-attention to capture global information for better co-speech gesture generation of arbitrary length depending on the speech duration. Furthermore, we exploit WavLM features \cite{DBLP:journals/jstsp/ChenWCWLCLKYXWZ22} to consider semantic, emotional, and other information in audio to improve the generalization and robustness of our model.
Finally, we use random masks to perform classifier-free guidance \cite{DBLP:journals/corr/abs-2207-12598} at training time and thus achieve the interpolation and editing of the control conditions.
The main contributions of our work are:
\begin{itemize}
    \item We extend the diffusion model with temporal information for audio-driven co-speech gesture generation. 
    By virtue of the diffusion model, we can have a high degree of control over the generated gestures, e.g., editing the style of the gestures, setting the initial gestures, and generating diverse gestures.
    \item We use cross-local attention and global self-attention to capture feature information and 
    make generated gestures that are more appropriate for speech.       
    \item Extensive experiments show that our model can generate human-like, speech-matched, style-matched gestures that significantly outperform existing gesture generation methods.
\end{itemize}

\section{Related Work}      

\subsection{Co-speech Gesture Generation}

Gesture generation is a complex task that requires understanding speech, gestures, and their relationships.
Data-driven approaches attempt to learn gesticulation skills from human demonstrations.
Present studies mainly consider four modalities: text \cite{DBLP:journals/corr/abs-2101-05684,DBLP:conf/icra/YoonKJLKL19},  
audio \cite{DBLP:conf/iva/HabibieXMLSPET21,9710107,DBLP:conf/cvpr/GinosarBKCOM19},         
gesture motion, and speaker identity \cite{yoon2020speech,DBLP:conf/cvpr/LiuWZXQLZWDZ22,DBLP:journals/cgf/AlexandersonHKB20}.         
\cite{DBLP:conf/iva/HabibieXMLSPET21} propose the first approach to jointly synthesize both the synchronous 3D conversational body and hand gestures, as well as 3D face and head animations.
\cite{yi2022generating} employ an autoencoder for face motions, and a compositional vector-quantized variational autoencoder (VQ-VAE) to generate more diverse gestures.
\cite{xie2022vector} introduce a VQ-VAE model to represent a pose sequence as a sequence of latent codes and develop a diffusion architecture for Text-to-Sign pose sequences generation.
As for learning individual styles \cite{DBLP:conf/iccv/0071KPZZ0B21,DBLP:conf/cvpr/LiangFZHP022}, 
\cite{yoon2020speech} propose the first end-to-end method for generating co-speech gestures using the tri-modality of text, audio and speaker identity.
\cite{DBLP:conf/eccv/AhujaLNM20} train a single model for multiple speakers while learning the style embeddings for the gestures of each speaker.
\cite{DBLP:conf/cvpr/LiangFZHP022} propose a semantic energized generation method for semantic-aware gesture generation.
\cite{DBLP:journals/tog/AoGLCL22} disentangle both low-level and high-level embeddings of speech and motion based on linguistic theory.

Some works use motion matching 
methods to generate co-speech gestures \cite{DBLP:conf/siggraph/HabibieESANNT22,DBLP:conf/icmi/ZhouBC22}.     
The approach requires careful design of the database, which is directly related to the performance of the generated gestures.
The length of matching needs to be balanced between quality and diversity. 
Furthermore, the approach also requires complex and time-consuming manual design of the matching rules.

Recently, a high-quality 3D gestures dataset ZeroEGGS \cite{ghorbani2022zeroeggs} is built from multi-camera videos with style labels. This dataset is used in our work.

\subsection{Diffusion Models for Motion Generation}

Diffusion models excel at modeling complicated data distribution and generating vivid motion sequences.
Many works integrate diffusion-based generative models into the motion domain and carefully adapt the network structure of classifier-free diffusion generative models for the human motion domain, such as based on Transformer \cite{kim2022flame,tevet2022human,ren2022diffusion,zhou2022ude}.       
 \cite{chang2022unifying} design a multi-task architecture of diffusion model and use adversarial and physical regulations for human motion synthesis.
\cite{DBLP:journals/corr/abs-2212-04495} introduce MoFusion to generate long, temporally plausible, and semantically accurate motions.
 A physics-guided motion diffusion model \cite{yuan2022physdiff} incorporates physical constraints into the diffusion process.

\cite{DBLP:conf/cvpr/GinosarBKCOM19} propose a cross-modal translation method based on the speech-driven gestures of a single speaker.
\cite{DBLP:conf/cvpr/SiyaoYGLW0L022} propose a cross-conditional causal attention layer to keep the coherence of the generated body.
\cite{10.1145/3536221.3558060} use locality-constraint attention mechanism and achieve the best gesture-speech appropriateness in the full-body level of the GENEA 2022 gesture generation challenge \cite{DBLP:conf/icmi/YoonWKVNTH22}.
MotionDiffuse \cite{zhang2022motiondiffuse} using cross attention enables probabilistic mapping, realistic synthesis, and multi-level manipulation.
In our work, we use cross-local attention to capture local information of gestures and speech; and then use self-attention to capture global information to make the generated gestures match better with the speech.

\section{Our Approach}      

\subsection{Diffusion Model for Gesture Generation}

Our idea is to generate gestures with a diffusion model \cite{ho2020denoising} by learning to gradually denoising starting from pure noise.
As shown in Figure \ref{fig:Framework}, the diffusion model consists of two parts: the forward process (diffusion process) $q$ and the reverse process (denoising process) $p_\theta$.

\textbf{Diffusion Process}.
The diffusion process $q$ is modeled as a Markov noising process.
We denote the generated gesture as $x$, which has the same dimension as an observation data $x_0 \sim q\left(x_0\right)$, $q\left(x_0\right)$ denotes the distribution of the real data.
According to a variance schedule $\beta_1, \beta_2, \ldots, \beta_T$ ($0<\beta_1<\beta_2<\cdots<\beta_T<1$, $T$ is the total time step),
we add Gaussian noise 
\begin{equation}
\label{Gaussian}
q\left(x_t \mid x_{t-1}\right)=\mathcal{N}\left(x_t ; \sqrt{1-\beta_t} x_{t-1}, \beta_t \mathbf{I}\right)
\end{equation}
to the gesture at each time $t$ gradually, and if the schedule is properly designed and $T$ is large, pure noise 
\begin{equation}
\label{eq1}
q\left({x}_{1: T} \mid {x}_0\right)=\prod_{t=1}^T q\left({x}_t \mid {x}_{t-1}\right)
\end{equation}
will be obtained at the end.

\textbf{Denoising Process}.
The denoising process $p_\theta$ is a process of learning parameter $\theta$ via a neural network.
Assuming that the denoising process also conforms to a Gaussian distribution, i.e., the noise $x_t$ at time $t$ is used to learn $\mu_\theta$, $\Sigma_\theta$, then 
\begin{equation}
p_\theta\left({x}_{t-1} \mid {x}_t\right)=\mathcal{N}\left({x}_{t-1} ; {\mu}_\theta\left({x}_t, t\right), {\Sigma}_\theta\left({x}_t, t\right)\right)
\end{equation}
For calculation convenience, we assume that $\alpha_t=1-\beta_t$ and $\bar{\alpha}_t=\prod_{i=1}^T \alpha_i$.
Then the noisy gesture $x_t$ at time $t$ can be written as 
\begin{equation}
q\left({x}_t \mid {x}_0\right)=\mathcal{N}\left({x}_t ; \sqrt{\bar{\alpha}_t} {x}_0,\left(1-\bar{\alpha}_t\right) \mathbf{I}\right)
\end{equation}
The network is optimized by minimizing the difference between the real noise $\epsilon$ and the predicted noise $\epsilon_\theta\left({x}_t, t\right)$ \cite{ho2020denoising}.
When sampling, we can learn the mean 
$\mu_\theta\left({x}_t, t\right)=\frac{1}{\sqrt{\alpha_t}}\left({x}_t-\frac{\beta_t}{\sqrt{1-\bar{\alpha}_t}} \epsilon_\theta\left({x}_t, t\right)\right)$ by fix the variance.

\textbf{Framework.}
Our goal is to synthesize a human gesture $x^{1:N}$ of length $N$ given conditions $c$.     
In our work, we follow \cite{ramesh2022hierarchical,tevet2022human} to predict the signal itself instead of predicting $\epsilon_\theta\left({x}_t, t\right)$ \cite{ho2020denoising}.
The Denoising module reconstructs the original signal $x_0$ based on the input noise $x_t$, noising step $t$ and conditions $c$
\begin{equation}
\hat{x}_0=\operatorname{Denoise}\left(x_t, t, c\right)
\end{equation}

Then the Denoising module can be trained by optimizing the Huber loss \cite{huber1992robust} between the generated poses $\hat{x}_0$ and the ground truth human gestures $x_0$ on the training examples:
\begin{equation}
\mathcal{L}=E_{x_0 \sim q\left(x_0 \mid c\right), t \sim[1, T]}\left[\operatorname{HuberLoss}(x_0-\hat{x}_0)\right]
\end{equation}      

\subsection{Attention-based Speech-driven Gesture Generation Model}

\textbf{Feature Processing in Denoising module.}
As shown in Figure \ref{fig:Framework}, gestures are generated based on noising step $t$, noisy gesture $x_t$ and conditions $c$ (including audio $a$, style $s$, and seed gesture $d$).
For each feature, the processing pipeline is as follows:
\begin{itemize}
    \item Noising step: During training, noising step $t$ is sampled from a uniform distribution of $\{1, 2, \dots, T\}$, with the same position encoding as \cite{vaswani2017attention}, and then mapped to a space $\textbf{T}$ of dimension 256 by a multilayer perceptron (MLP).
    \item Noisy gesture: During training, $x_t$ is the noisy gesture with the same dimension as the real gesture $x_0$ obtained by sampling from the standard normal distribution $\mathcal{N}(0,\mathbf{I})$. 
    When sampling, the initial noisy gesture $x_T$ is sampled from the standard normal distribution and the other $x_t, t<T$ is the result of the previous noising step.
    Then the dimension is adjusted to 256 as $\textbf{G}$ by a linear layer.
    \item Audio: All audio recordings are downsampled to 16kHz, and features are generated from the pre-trained models of WavLM Large \cite{DBLP:journals/jstsp/ChenWCWLCLKYXWZ22}.
    We use linear interpolation to align WavLM features and gesture $x_0$ in the time dimension to 20fps, and then use a linear layer to reduce the dimension of features to 64 forming the final audio feature sequence $\textbf{A}$.
    \item Style: The styles of gestures are represented as one-hot vectors where only one element of a selected style is nonzero, mapping to the 64-dimensional space $\textbf{S}$ via a linear layer.
    \item Seed gesture: Seed gesture helps to make smooth transitions between consecutive syntheses \cite{yoon2020speech}. 
    Please see our supplementary material for more detail regarding the ground truth gestures clip $\textsl{g}\in \mathbb{R}^{(8+N)\times1141}$.
    The first 8 frames of the gestures clip $\textsl{g}$ are used as the seed gesture $d$ and the remaining $N$ frames are used as the real gesture $x_0$ to calculate loss $\mathcal{L}$.
    Then we map the feature dimensions of seed gesture $d$ to the space $\textbf{D}$ of 192 dimensions using a linear layer.
    The length of our generated gesture is 4 seconds and we resample the gesture animation to 20 fps, then $N$ = 80.

\end{itemize}

\textbf{Model in Denoising module.}
We implement denoising with the attention-base architecture.
Before utilizing long-range correlations, it is advisable to build up representations with local context \cite{DBLP:conf/acl/RaeR20}.

\begin{figure}[!t]
  \centering
  \subfigure[full self-attention]{
  \label{Fig.sub.1}
  \includegraphics[width=0.23\linewidth]{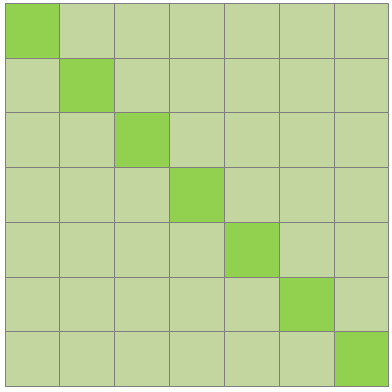}}
  \enspace
  \subfigure[sliding window attention]{
  \label{Fig.sub.2}
		\includegraphics[width=0.23\linewidth]{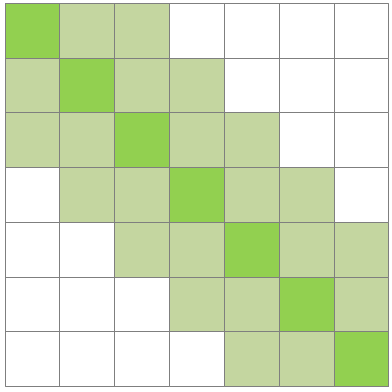}}
  \enspace
  \subfigure[cross-local attention]{
  \label{Fig.sub.3}
		\includegraphics[width=0.23\linewidth]{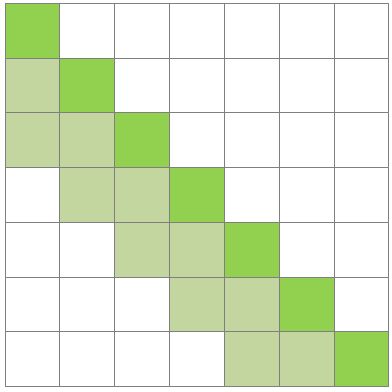}}
  \caption{Different patterns of attention used in our experiments, where (a) and (c) are attention mechanisms used in our model and (b) is a pattern compared in Section \ref{Ablation_sec}. 
  The rows represent the outputs and the columns represent the inputs. The colored squares highlight the relevant elements for each row of output.
  }
  \label{attention}
\end{figure}

We concatenate the seed gesture $\textbf{D}$ and style $\textbf{S}$ together to form a 256-dimensional vector, and then add the information of the noising step $\textbf{T}$ to form $\textbf{Z}$.
Our network takes the vector $\textbf{Z}$ and stacks its replicates into a sequence feature to align with the timeline of audio and gesture features, which are then concatenated with the audio $\textbf{A}$ and gesture $\textbf{G}$ as the input to the cross-local attention network.
Our proposed cross-local attention for co-speech gesture generation is based on Routing Transformer \cite{DBLP:journals/tacl/RoySVG21}, which shows that local attention is important in building intermediate representations, as shown in Figure \ref{Fig.sub.3}.

After that, we concatenate the output of cross-local attention with $\textbf{Z}$ and feed it into the self-attention network, as shown in Figure \ref{Fig.sub.1}.
The self-attention mechanism is similar to Transformer \cite{vaswani2017attention} encoder, which determines the computational dependencies between the sequential elements of the data, and is implemented as
\begin{equation}
\operatorname{Attention}(\mathbf{Q}, \mathbf{K}, \mathbf{V}, \mathbf{M})=\operatorname{softmax}\left(\frac{\mathbf{Q} \mathbf{K}^T+\mathbf{M}}{\sqrt{C}}\right) \mathbf{V}
\end{equation}
where $\mathbf{Q}$, $\mathbf{K}$, $\mathbf{V}$ denote the query, key and value from input,
and $\mathbf{M}$ is the mask, which determines the type of attention patterns. 
We use the same relative position encoding (RPE) mechanism as \cite{DBLP:conf/iclr/KitaevKL20} so that the temporal effect on gesture translation is invariant.
Finally, the output of self-attention is mapped back to the same dimension as $x_0$ after a linear layer.

From Figure \ref{attention}, we can find these different attention mechanisms can be achieved by simply adjusting the corresponding mask $\mathbf{M}$, we also experimented with sliding window attention in Longformer \cite{DBLP:journals/corr/abs-2004-05150}, the results are analyzed in Section \ref{Ablation_sec}.

\textbf{Sample Module.}     
The final co-speech gesture is given by splicing a number of clips of length $N$.
The seed gesture for the first clip can be generated by randomly sampling a gesture from the dataset or by setting it to the average gesture. Then the seed gesture for other clips is the last 8 frames of the gesture generated in the previous clip.
For every clip, in every noising step $t$, we predict the clean gesture $\hat{x}_0$ =$\operatorname{Denoise}(x_t, t, c)$, and add the noise to the noising step $x_{t-1}$ using Equation (\ref{Gaussian}) with the diffuse process.
This process is repeated from $t$ = $T$ until $x_0$ is reached (Figure \ref{fig:Framework} bottom).

\begin{figure*}[!ht]
  \centering
  \subfigure[Box plot of ratings in human-likeness.]{
  \label{Fig.box}
  \includegraphics[width=0.31\linewidth]{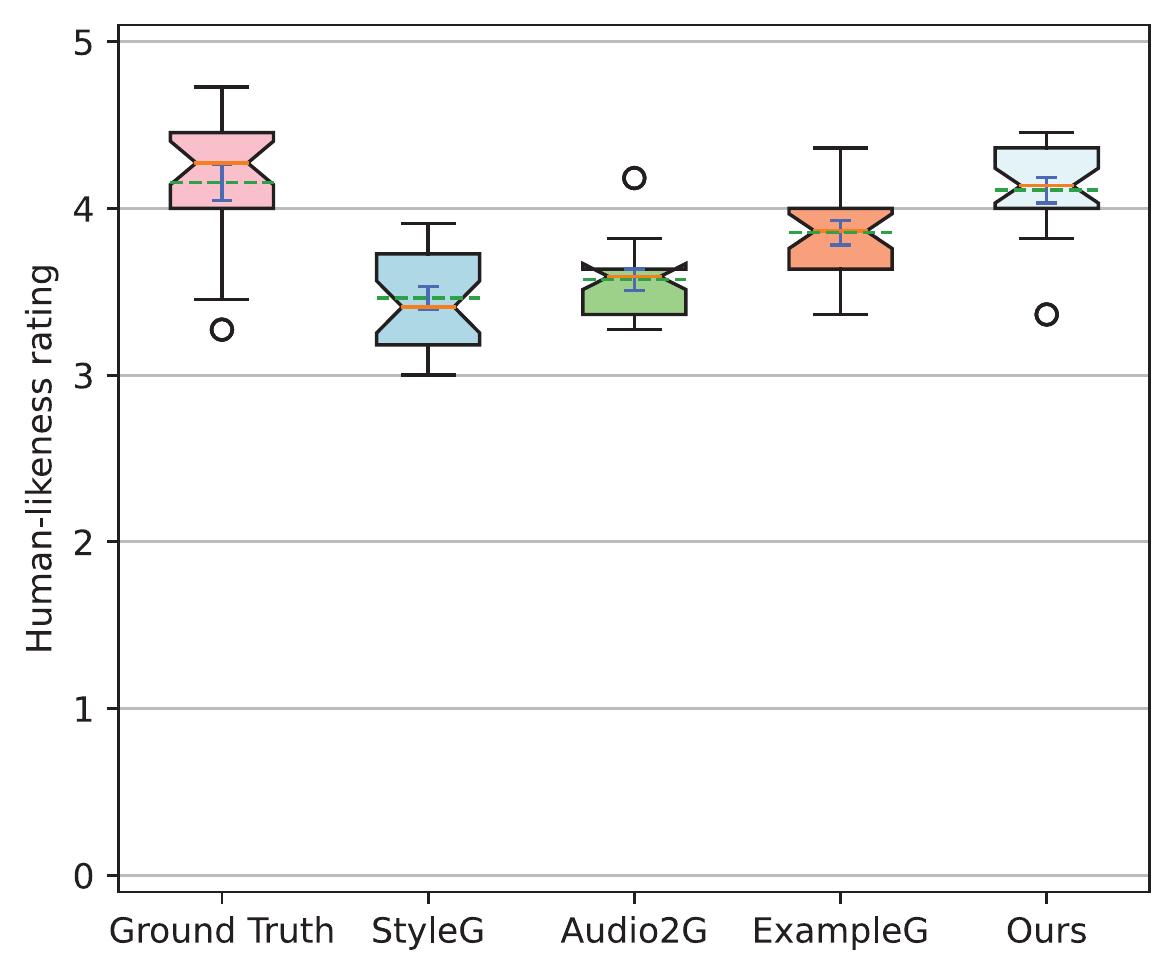}}
  \enspace
  \subfigure[Box plot of ratings in gesture-speech appropriateness.]{
		\includegraphics[width=0.31\linewidth]{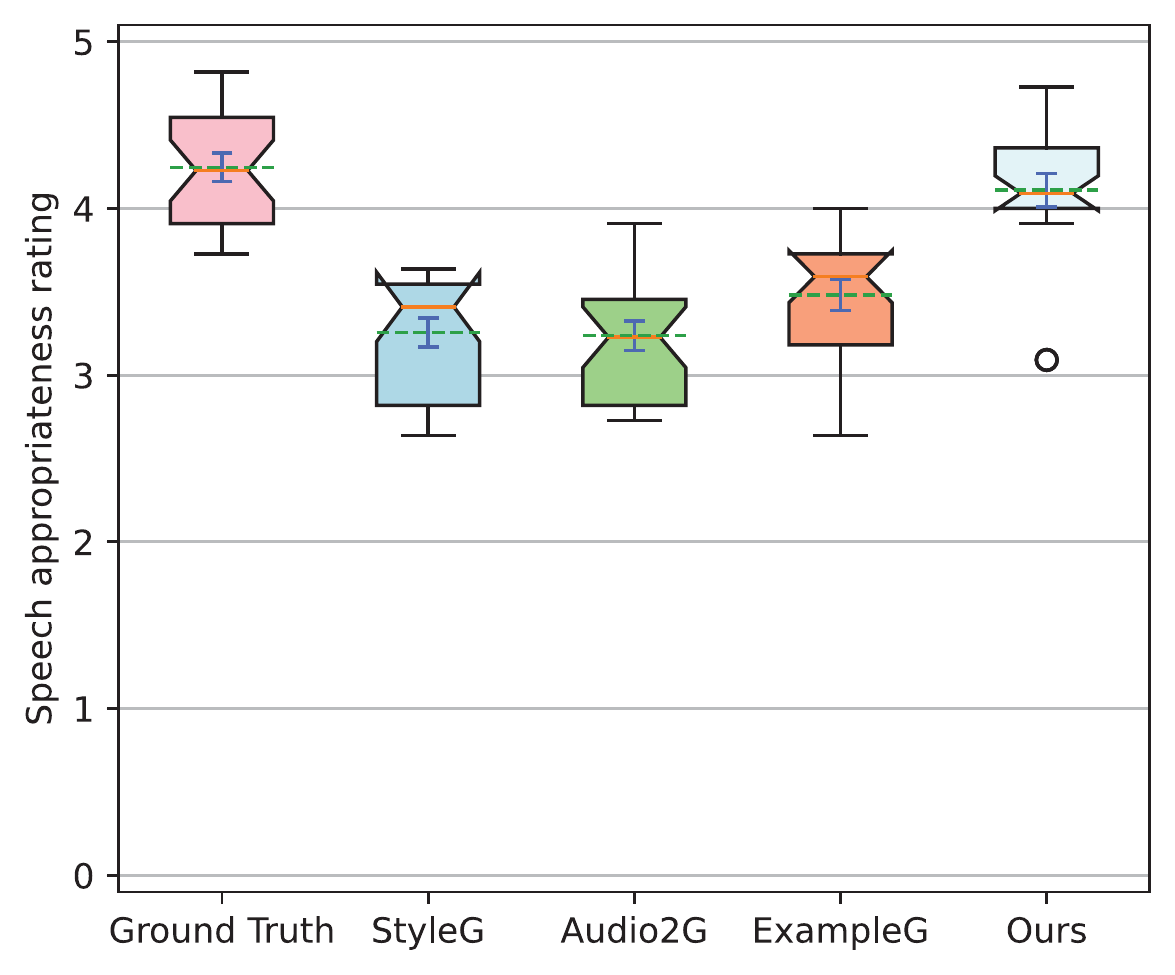}}
  \enspace
  \subfigure[Box plot of ratings in gesture-style appropriateness.]{
		\includegraphics[width=0.31\linewidth]{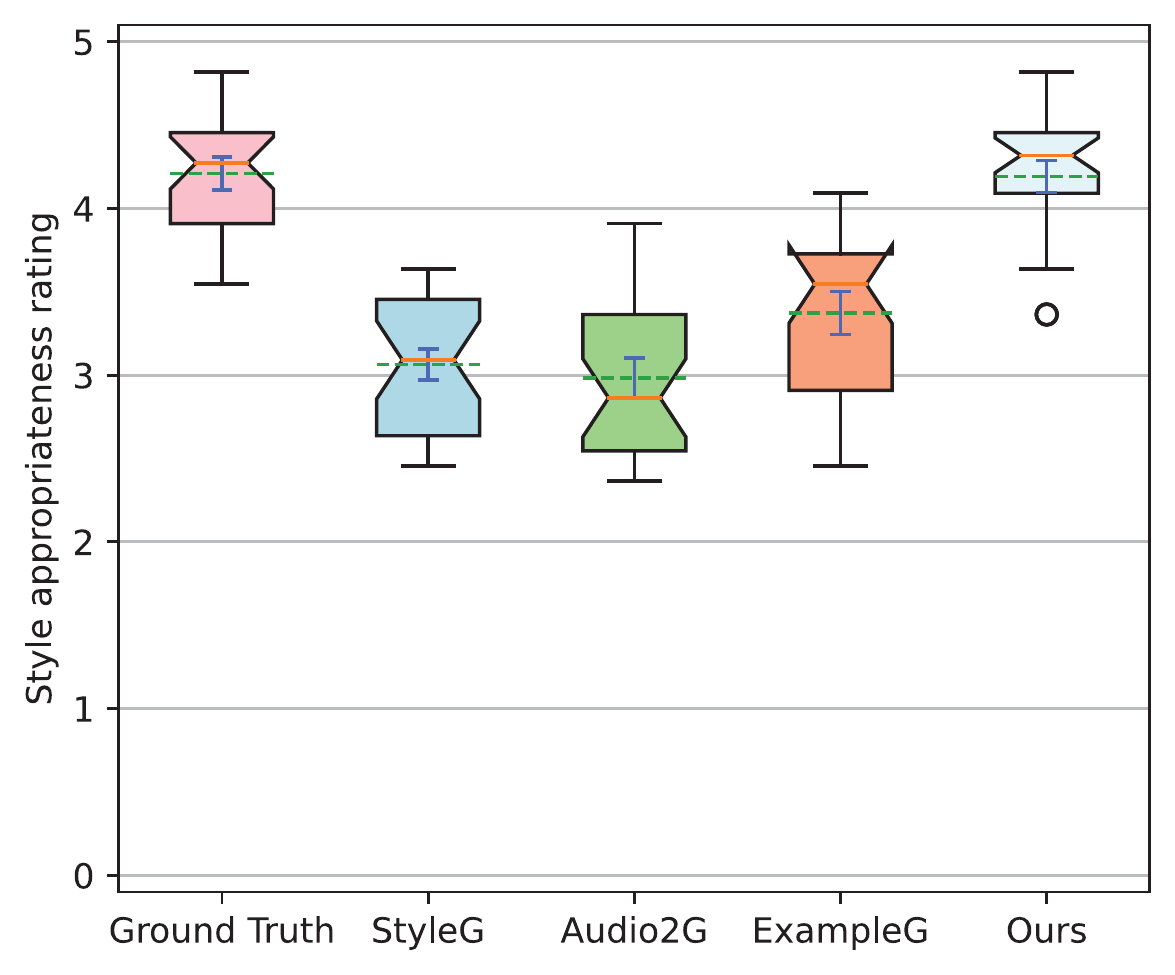}}
  \caption{
  Box plot visualizing comparison results of MOS for different models in different dimensions.
 The box extends from the first lower quartile (Q1) to the third greater quartile (Q3) of the data.
 The red line denotes the median.
 The notches represent the 95\% confidence interval (CI) around the median.
 When the CI is less than Q1 or greater than Q3, the notch extends beyond the box, giving it a unique ``flipped'' appearance.
 We have also marked the mean and its 95\% CI in the figure with a green dashed line and a blue vertical line, respectively.}
  \label{mos}
\end{figure*}

\subsection{Style-controllable Gesture Generation}

Since the algorithm generates gestures based on control conditions, the control conditions can be not only audio $a$ but also style $s$, or seed gesture $d$, etc.
As shown in Figure \ref{fig:Framework}, we refer to \cite{DBLP:journals/corr/abs-2207-12598,tevet2022human} and add random masks to the pipeline of seed gesture $d$ and style $s$ feature processing for classifier-free learning, which enables accurate control of different conditions.
Then, the classifier-free guidance of gestures generation can be achieved by combining the predictions of the conditional model $\operatorname{Denoise}\left(x_t, t, c_1\right), c_1 = [d, s, a]$ and the unconditional model $\operatorname{Denoise}\left(x_t, t, c_2\right), c_2 = [\varnothing, \varnothing, a]$ during the training process, as Equation (\ref{8}).
\begin{align}
\label{8}
\hat{x}_{0{\gamma},{c_1},{c_2}}&=\gamma\operatorname{Denoise}\left(x_t, t, c_1\right) + (1-\gamma)\operatorname{Denoise}\left(x_t, t, c_2\right)
\end{align}

In practice, the Denoising module learns both the conditioned and the unconditioned distributions by randomly masking 10\% of the samples using Bernoulli masks.
Then, 
as for style $s$ in condition, we can generate style-controlled gestures when sampling by interpolating or even extrapolating the two variants using $\gamma$, as $c_1 = [d, s_1, a], c_2 = [d, s_2, a]$ in Equation (\ref{8}).
Please refer to our supplementary material for training details such as dataset and implementation details.



\section{Experiments}       

\subsection{Comparison to Existing Methods}

We compare our proposed model with StyleGestures \cite{https://doi.org/10.1111/cgf.13946}, Audio2Gestures \cite{9710107}, ExampleGestures \cite{ghorbani2022zeroeggs}.
Currently, speech-driven gestures lack objective metrics that are consistent with human subjective perception \cite{DBLP:conf/icmi/YoonWKVNTH22,DBLP:conf/iui/KucherenkoJYWH21,DBLP:journals/corr/abs-2211-09707}, even for Fr\'{e}chet gesture distance (FGD) \cite{yoon2020speech,DBLP:journals/corr/abs-2212-04495}, so all our experimental scoring are done by human subjective evaluation.
We conduct the evaluation on three dimensions. 
The first two follow the evaluation in GENEA \cite{DBLP:conf/icmi/YoonWKVNTH22}, which evaluates human-likeness and gesture-speech appropriateness. 
The third dimension is gesture-style appropriateness.

\textbf{User Study.} 
To understand the real visual performance of our method, we conduct a user study among the gesture sequences generated by each compared method and the ground truth motion capture data. 
The length of the evaluated clips ranged from 11 to 51 seconds, with an average length of 31.6 seconds.
Note that the clip gestures used for the subjective evaluation here are longer compared to the GENEA \cite{DBLP:conf/icmi/YoonWKVNTH22} evaluation (8-10 seconds), as a longer period time could produce more pronounced and convincing appropriateness results \cite{DBLP:conf/icmi/Yang0LZLCB22}.
Participants rated at a 1-point interval from 5 to 1, with labels (from best to worst) of ``excellent", ``good", ``fair", ``poor", and ``bad".
More details about the user study are shown in the supplementary material.



The mean opinion scores (MOS) on human-likeness, speech appropriateness, and style appropriateness are reported in Figure \ref{mos}.
If the notches of the two boxes do not overlap, we can consider this as strong evidence that the distributions are significantly different \cite{mcgill1978variations}.
Our method significantly surpasses the compared state-of-the-art methods with human-likeness, gesture-speech, and gesture-style appropriateness, and even produces competitive results with ground truth in all three dimensions.
According to the feedback from participants, our generated gestures are ``more semantically relevant", ``more natural", and ``match the style", while our approach has ``foot-sliding" compared to Ground Truth.
However, this is a common problem for non-physical-based motion generation systems and could be solved by post-processing \cite{ghorbani2022zeroeggs,DBLP:journals/corr/abs-2301-05175}.

\subsection{Gesture Controllability}

\begin{figure}[!t]
  \centering
   \includegraphics[width=\linewidth]{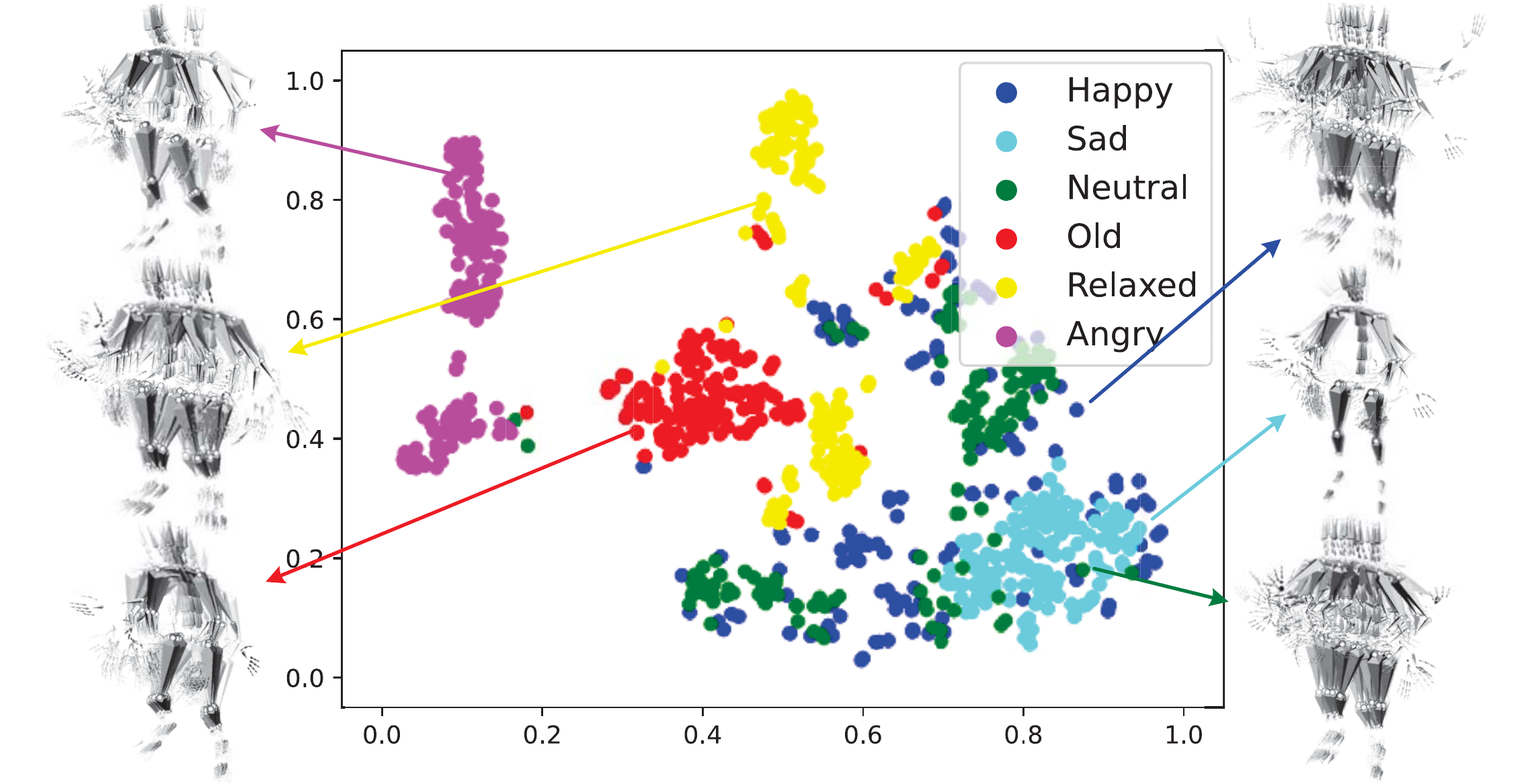}
   \caption{The tSNE visualization of gestures with different styles and the shadow maps of the skeletal gesture with the corresponding style.      
   For example, for the `old' gesture, its waist and knees are more bent, and its hands are basically on the knees or waist.}
   \label{fig:Style-PCA}
\end{figure}

\textbf{Style Control.} 
Assuming that the neutral audio does not affect the gesture style, then we can generate stylized gestures with a neutral speech by setting $\gamma=1$ and $s$ in equation (\ref{8}). 
We choose two speech segments in the test set with neutral audio to generate six stylized gestures respectively. 
Figure \ref{fig:Style-PCA} illustrates generated gesture $\hat{x}_t$ of different input style $s$ with the same neutral audio visualized by the tSNE method.

We also plotted the body skeleton generated by the corresponding style in the figure, and it can be noticed that for the `old' style, its waist and knees are more bent, and its hands are basically on the knees or waist; for the `sad' style, its head is hanging and its hands are in a lower position; for the `relaxed' style, its hips are forward and its standing posture is relaxed; for the `angry' style, its hands move up and down quickly.
Note that although the differences between the `neutral' and `happy' style gestures are still relatively obvious in the stylistic visualization of the skeleton, i.e., for the `happy' style, its hand position is higher and its amplitude is larger,
their tSNE is almost coupled together. 
In our analysis, this happens because the so-called neutral speech still contains information such as emotions and semantics that is contained in the WavLM features.
The balance of the style from audio $a$ and from style $s$ can be further controlled by the editing the style intensity $\gamma$.

\begin{figure}[!t]
  \centering
   \includegraphics[width=0.84\linewidth]{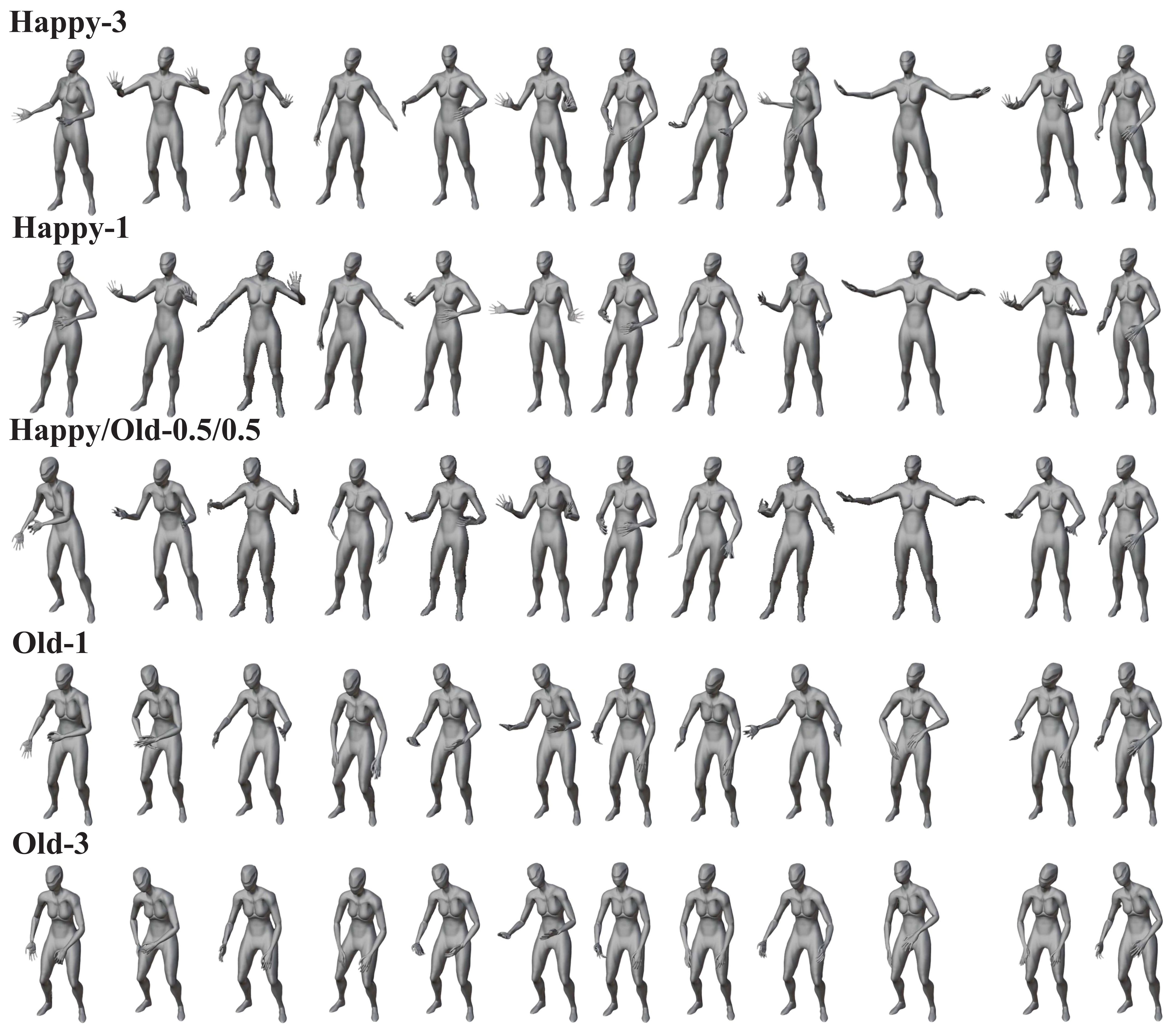}
   \caption{Style editing and interpolation results.        
   From top to bottom, the body twists and hand movements gradually decrease and the hand position becomes lower. 
   Despite the change in style, the generated gestures still match well with speech in different styles.}
   \label{fig:strength}
\end{figure}

\textbf{Style Edit.}
To further analyze the relationship between style intensity and the style implied by the speech, we choose the `happy' style and the `old' style and set $\gamma$ = 1 and 3 in equation \ref{8}. 
Also, to compare the results, we set $\gamma$ = 0.5 in equation \ref{8}, in order to interpolate the different styles.
The other parameters remain unchanged, and we plot a 12-second, FPS = 1 gesture generation result as shown in Figure \ref{fig:strength}.

As shown in Figure \ref{fig:strength}, we can see that when we use the `happy' style with $\gamma$ = 3, both its body rotation and hand movements are the largest, and its hand positions are the highest; in contrast, when we use the `old' style with $\gamma$ = 3, its waist is the most bent, the hands are barely lifted, and there is no much change in the whole movement sequence;
As for the other three results, 
the intensities of their styles are in the middle of the above two, 
and the style gradually changes from happy to old from top to bottom.
Due to our model architecture, 
the generated gestures and speech are more appropriate, 
even though these styles are not the same.
Notice that when we use `happy' and `old' styles with $\gamma$ = 0.5, the result is closer to using the `happy' style with $\gamma$ = 1, while the `old' style is almost imperceptible.
This observation further validates the previous finding that the `happy' style is embedded in the `neutral' speech used for testing.
This finding is useful, for example, we found in our experiments that if we want to control the `happy' speech to generate the `sad' gesture, $\gamma$ = 1 is basically ineffective because the model can learn the happy style from the speech.
Since there is such a coupling between the style of speech and the style of gestures, then setting a larger $\gamma$ can edit the style better.
Thus we are able to generate gestures that do not exist in the original dataset (e.g., gestures for `happy' speech in the `sad' style) by style intensity.

\begin{figure}[!t]
  \centering
\includegraphics[width=0.875\linewidth]{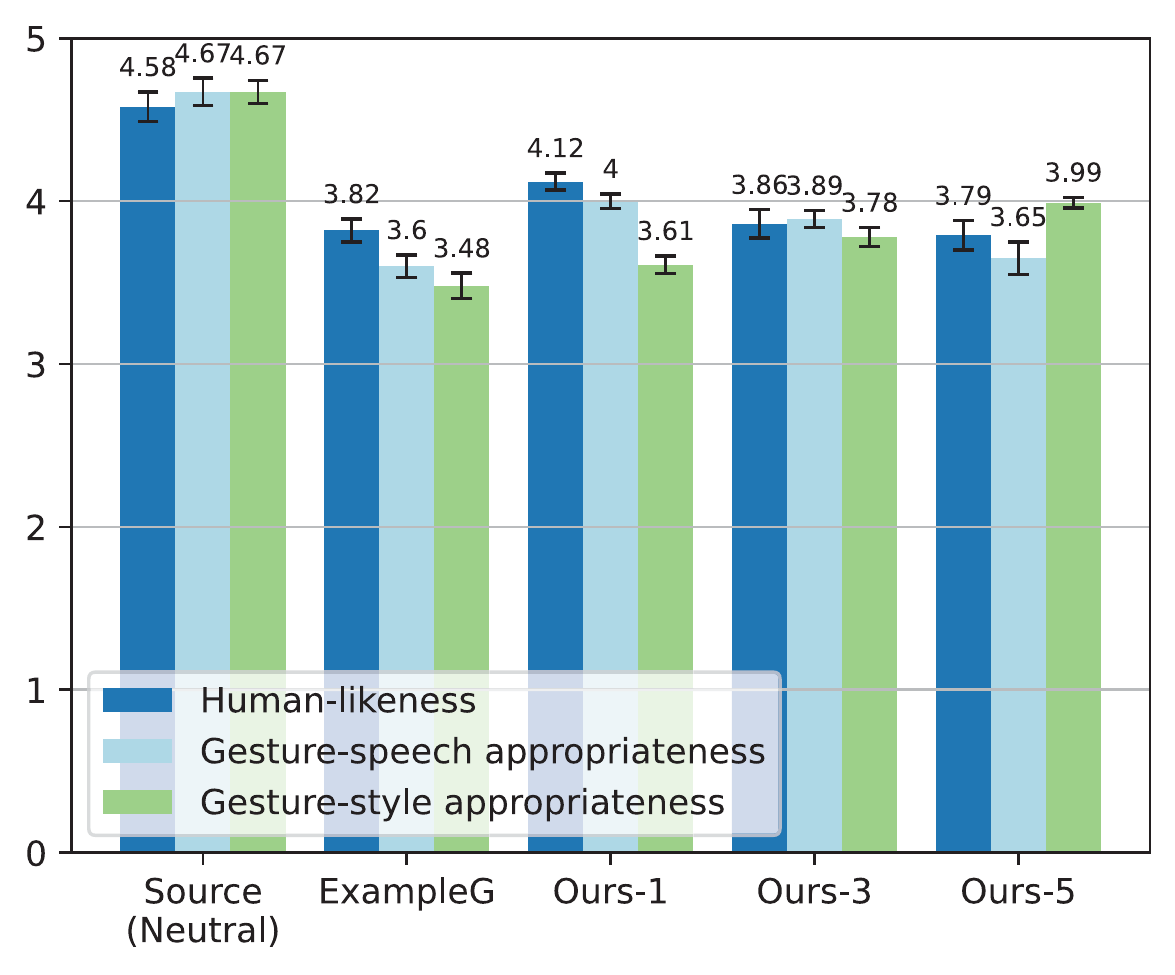}
   \caption{Average results of MOS with 95\% confidence intervals for three dimensions.
   'Ours-$\gamma$' denotes the style control intensity $\gamma$ of our model.
   Our model significantly outperform ExampleGesture 
   overall and could editing the intensity of the styles.
   The parameter $\gamma$ increases and the other two scores will reasonably decrease.
   }
   \label{fig:4}
\end{figure}

\textbf{User Study.}
Further, we would like to explore the relationship between style intensity and human-likeness and speech appropriateness, so we conducted a user study.
To avoid styles in speech from influencing participants' scoring, as before, we only control the intensity of the styles for a neutral speech and then asked participants to score the three previous dimensions.
ExampleGesture \cite{ghorbani2022zeroeggs} can also control the generation of different styles of gestures from the same speech. Hence we choose it as the baseline model. Since the gestures generated here do not exist in the dataset, the source neutral speech with neutral style is used as a reference.
The results are shown in Figure \ref{fig:4}.

The results show that our model is similar to ExampleGesture in terms of gesture-style appropriateness of the results at $\gamma$ = 1, and our human-likeness and speech appropriateness significantly exceed ExampleGesture.
Meanwhile, the style is significantly more appropriate when $\gamma$ increases, but the scores of the other two dimensions decrease. 
This is also intuitive, i.e., if the intensity of the `old' style is too high, the hands are barely lifted and the entire motion sequence are small in amplitude, so it looks less human-like and less appropriate to the speech.
We also find that the results of generating style control (Figure \ref{fig:4}) were degraded compared to the results of directly generating the style corresponding to the speech (Figure \ref{mos}). We believe that controlling one style of speech to generate another style of gesture is in itself a ``difficult and conflicting" task because the style of speech and the style of gesture are still related and coupled together.

\begin{figure}[!t]
  \centering
   \includegraphics[width=\linewidth]{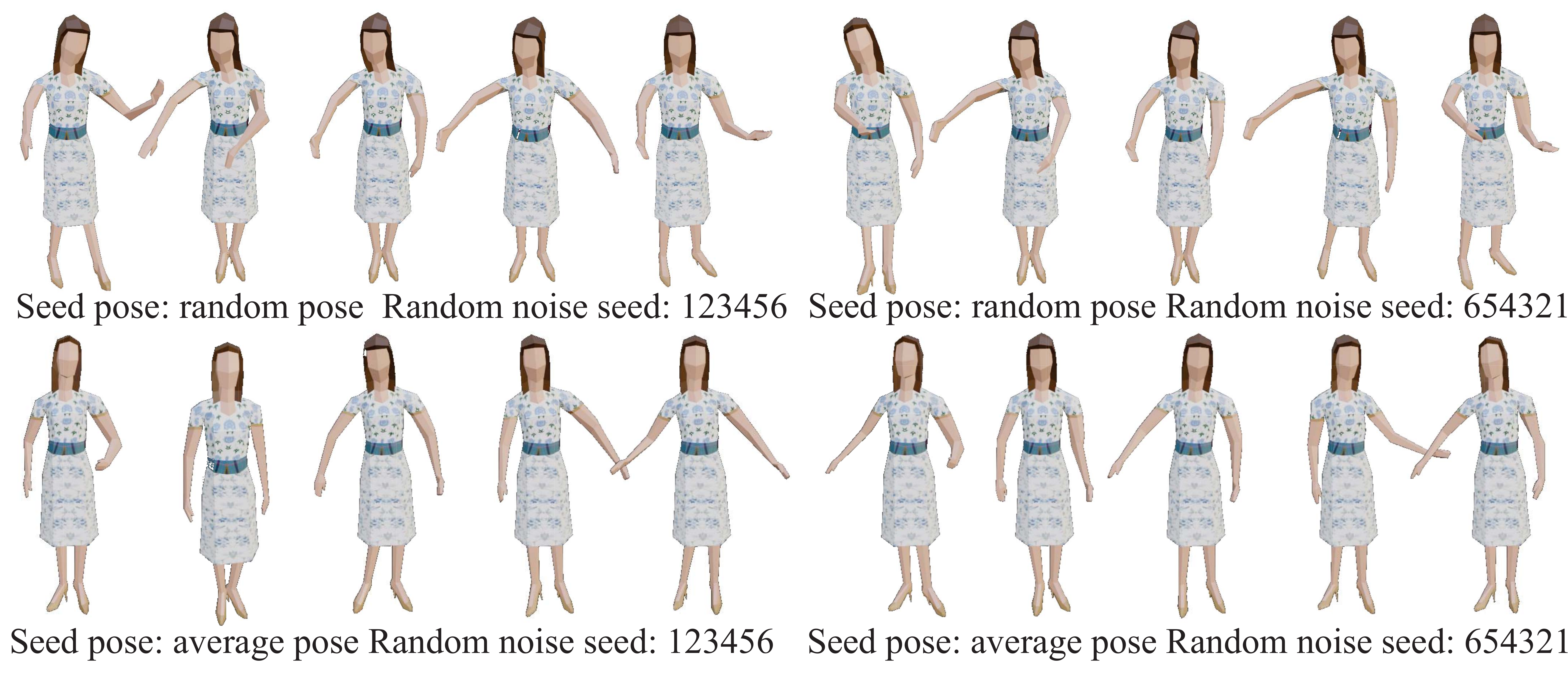}
   \caption{Visualization of the diversity of generated gestures. People make different co-speech gestures at different moments in different states. Just like real people, for the same speech, our method is able to generate different gestures with different seed gesture or with different noisy gesture.}
   \label{fig:seed}
\end{figure}

\textbf{Generate diverse gestures.}
Due to our model architecture, even for the same speech and style, different noisy gesture and different seed gesture could generate different results, as shown in Figure \ref{fig:seed}. This is the same as real human speech, which creates diverse co-speech gestures related to the initial position.
Our analysis before was performed on the style dimension.
Note that the model also adds a random mask to the processing of the seed gesture, 
so it can also interpolate and extrapolate different seed gestures to control the generation of different and diverse initial position gesture.

\subsection{Ablation Studies}
\label{Ablation_sec}

Moreover, we conduct ablation studies to address the performance effects of different components in our model. 
Since gesture-style appropriateness can be controlled by parameter and affect the other two dimensions, we set $\gamma$ to 1 and score only human-likeness and speech appropriateness for ease of comparison.
The results of our ablations studies are summarized in Table \ref{Ab}. The visual comparisons of this study can be also referred to the supplementary video. 
We explore the effectiveness of the following components: (1) WavLM features (2) local attention (3) local attention pattern (4) self-attention (5) attention.
We conduct the experiments on each of the five components, respectively.

\begin{table}[!t]
\footnotesize
\centering
\resizebox{\columnwidth}{!}{%
\begin{tabular}{lcc}
\hline
\multicolumn{1}{c}{Name} &
  \begin{tabular}[c]{@{}c@{}}Human\\ likeness \end{tabular}$\uparrow$ &
  \begin{tabular}[c]{@{}c@{}}Gesture-speech\\ appropriateness\end{tabular}$\uparrow$ \\ \hline
Ground Truth          & 4.15 $\pm$ 0.11          & 4.25 $\pm$ 0.09          \\
Ours                  & \textbf{4.11 $\pm$ 0.08} & \textbf{4.11 $\pm$ 0.10} \\
\quad$-$ WavLM             & 4.05 $\pm$ 0.10          & 3.91 $\pm$ 0.11          \\
\quad$-$ cross-local attention   & 3.76 $\pm$ 0.09          & 3.51 $\pm$ 0.15          \\
\quad$-$ self-attention    & 3.55 $\pm$ 0.13          & 3.08 $\pm$ 0.10          \\
\quad$-$ attention + GRU&
  3.10 $\pm$ 0.11 &
  2.98 $\pm$ 0.14 \\
\quad$+$ forward attention & 3.75 $\pm$ 0.15          & 3.23 $\pm$ 0.24          \\
 \hline
\end{tabular}%
}
\caption{Ablation studies results. '$-$' indicates modules that are not used and '$+$' indicates additional modules. Bold indicates the best metric.}
\label{Ab}
\end{table}

\textbf{User Study.}
Supported by the results in Table \ref{Ab}, when we do not use the WavLM feature but use the first 13 coefficients of the Mel-frequency cepstral coefficients (MFCC) instead, the scores of both dimensions decreased, especially the speech appropriateness. This is because the features extracted by the pre-trained WavLM model contain more information such as semantics and emotions, which is helpful to generate the corresponding gestures.
When there is no cross-local attention, the scores of both dimensions drop a lot. Because many gesture generation steps only involve short-range correlations, local attention can capture local information better, which is consistent with the observation of \cite{DBLP:conf/acl/RaeR20}.
Only self-attention relying on global information of long sequences becomes less effective.
Both human-likeness and gesture-speech appropriateness drop more when self-attention is removed, 
suggesting that self-attention is more important than local attention because there is inherent asynchrony in speech and gesture, and it is difficult to learn enough gestural information from only a local window (nearly half a second) of speech.
When attention is not used, we replace it with a GRU-based model, which has the worst results among all models, further illustrating the effectiveness of the attention mechanism.
In addition, we experiment using the attention structure in Figure \ref{Fig.sub.2} and find that the effect gets worse. 
The only difference between adding forward attention and the cross-local attention used in our model is that the gesture is generated with an extra look at the speech information in a future window. 
This is an exciting finding, although there is an inherent asynchrony between speech and gesture, in some ways it could indicate that gestures are more related to a small period of time in the present and the past and not to a short period of time in the future. 
In other words, people are more likely to say ``hello" before waving than to wave before saying ``hello".
It is also possible that different people have different styles and this dataset has only one actor that needs to be studied further.

\section{Discussion and Conclusion}

In this paper, we propose DiffuseStyleGesture, a diffusion model based method for audio-driven co-gesture generation. 
DiffuseStyleGesture demonstrates three major strengths: 
1) Based on a diffusion model, probabilistic mapping enhances diversity while enabling the generation of high-quality, human-like gestures. 
2) Our model synthesis gestures match the audio rhythm and text semantics based on cross-local and self-attention mechanisms.
3) Using the classifier-free guidance training approach, we can manipulate specific conditions, i.e., style and initial gesture, and perform interpolation or extrapolation to achieve a high degree of control over the generated gestures.
The subjective evaluation shows that our model outperforms existing arts on the temporal task of audio-driven co-gesture generation and demonstrates superior style manipulation.
There is room for improvement in this research, for example, solving the problem of many sampling steps and long-time consumption of diffusion methods for use in real-time systems is our future research direction.

\section*{Acknowledgments}

This work is supported by National Natural Science Foundation of China (62076144), Shenzhen Science and Technology Program (WDZC20220816140515001, JCYJ20220818101014030) and Shenzhen Key Laboratory of next generation interactive media innovative technology (ZDSYS20210623092001004).

\clearpage


\bibliographystyle{named}
\bibliography{main}

\renewcommand\thesection{\Alph{section}}
\setcounter{section}{0}
\section{Overview}
 In this supplementary file, we present more experimental results analysis.
\begin{itemize}
\item We describe the features process pipeline of the ground truth gesture clip in detail.
\item We give training details, including dataset descriptions and implementation details.
\item We show details of the user study we conduct.
\end{itemize}
\section{Groud Truth Gesture Clip}

Each frame of the reference gesture clip is represented by a feature vector $\mathbf{g}$.
Each frame of the reference gesture clip is represented by a feature vector $\mathbf{g}=\left[\mathbf{r}_p, \mathbf{r}_r, \dot{\mathbf{r}}_p, \dot{\mathbf{r}}_r, \rho_p, \rho_r, \dot{\rho}_p, \dot{\rho}_r, g_d\right]$ where $\mathbf{r}_p \in \mathbb{R}^{3}$ and $\mathbf{r}_r\in \mathbb{R}^{4}$ are root positions and rotations, $\dot{\mathbf{r}}_p\in \mathbb{R}^{3}$ and $\dot{\mathbf{r}}_r\in \mathbb{R}^{3}$ are root positional and rotational velocity,      
     $\rho_p \in \mathbb{R}^{3 j}$ and $\rho_r \in \mathbb{R}^{6 j}$ are the local 3D joint positions and the local 6D joint rotations, $\dot{\rho}_p \in \mathbb{R}^{3 j}$ and $\dot{\rho}_r \in \mathbb{R}^{3 j}$ are  the local 3D joint positional velocities and the local 3D joint rotational velocities, $j$ is the number of joints, and $g_d\in\mathbb{R}^{3}$ is the direction of gaze.
The number of joints used here is 75, then $\mathbf{g}\in \mathbb{R}^{1141}$.
    The gestures clip $\textsl{g}\in \mathbb{R}^{(8+N)\times1141}$ used during denoising are fixed-length segments.

\section{Training Details}
\textbf{Dataset.} We perform the training and evaluation on the recent Zeroeggs dataset proposed in \cite{ghorbani2022zeroeggs}, which contains 67 sequences of monologues performed by
a female actor speaking in English and covers 19 different motion styles, and the total length of the dataset is 135 minutes.
We choose six more typical and longest-duration styles (happy, sad, neutral, old, relaxed, angry) for training and validation.
And we divided the data into 8:1:1 by training, validation, and testing.

\textbf{Implementation Details.} In this work, gesture data are cropped to a length of $N$ = 80 (4 seconds, 20 fps).
We apply standard normalization (zero mean and unit variant) to all joint feature dimensions, and use mirror enhancement and length ratio enhancement for gesture data.
The cross-local attention networks use 8 heads, 32 attention channels, 256 channels, the window size is 11 (about 0.5 second), each window looks at the one window before it, and with a dropout of 0.1.
As for self-attention networks are composed of 8 layers, 8 heads, 32 attention channels, 256 channels, and with a dropout of 0.1.
Using the AdamW \citeA{DBLP:conf/iclr/LoshchilovH19} optimizer (learning rate is 3$\times 10^{-5}$) with a batch size of 384 for 300000 samples. 
Our models have been trained with $T$ = 1000 noising steps and a cosine noise schedule.
The whole framework can be learned in about 3 days on one NVIDIA V100 GPU.

\begin{figure}[!ht]
  \centering
\includegraphics[width=0.45\textwidth,height=0.9\textwidth]{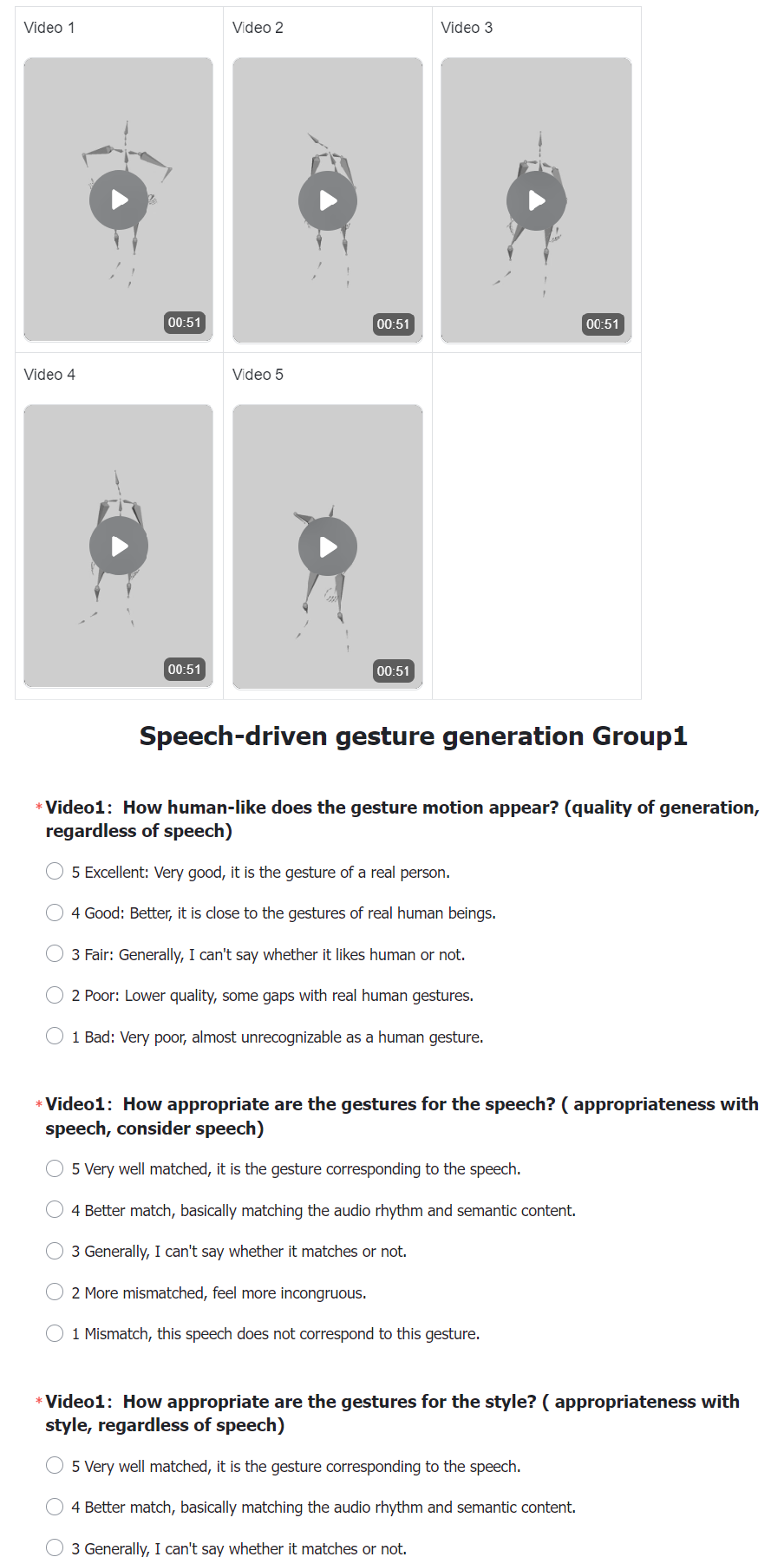}
  \caption{Screenshot of the parallel rating interface from the user study for comparison with existing methods.}
  \label{fig:comparison with existing methods}
\end{figure}

\section{User Study}

For each method, from the test set, we selected 10 segments of test speech and corresponding test motion to be used in the evaluations. 
We made sure there were no spoken phrases that ended on a “cliffhanger” in the evaluation. 
The experiment is conducted with 18 participants separately. 
The generated gesture data is visualized via Blender \footnote{https://www.blender.org/} rendering.

The experiment is conducted with participants with good English proficiency to evaluate the human-likeness, gesture-speech appropriateness and gesture-style appropriateness.
During the evaluation, we prompted the participants to ignore the finger movements.
They need to watch a video and then score the three dimensions.
The participants were almost all students from our lab \footnote{\url{https://thuhcsi.github.io/}} and a few from other schools, aged between 20 and 30.

For human-likeness evaluation, each evaluation page asked participants “How human-like does the gesture motion appear?” 
In terms of gesture-speech appropriateness evaluation, each evaluation page asked participants “How appropriate are the gestures for the speech?” 
As for gesture-style appropriateness evaluation, participants were asked “How appropriate are the gestures for the style?” 
Each page presented five videos of model results to be rated on a scale from 5 to 1 with 1-point intervals with labels (from best to worst) “Excellent”, “Good”, “Fair”, “Poor”, and “Bad”. 

An example of the evaluation interface for  different models in different dimensions can be seen in Figure \ref{fig:comparison with existing methods}.
Participants reported that the gestures generated by our framework contain many semantic and rhythmically related gestures.
Please refer to our supplementary video for more visualization results.

\section{More discussion}
About controllability. Since we use one-hot encoding to represent style, we only need to control the $\gamma$ in the equation (\ref{8}).
It is found that speech with style $\gamma$ of 1 has a more obvious corresponding style; natural speech or speech against anticipation (e.g. happy speech generating sad gestures) may need to set $\gamma$ to 2 or 3 will have an obvious style.
This is because speech and style themselves are coupled together.

\bibliographystyleA{named}
\bibliographyA{main.bib}

\end{document}